# Automated Comment Moderation Enhances Social Media Advertising Performance

Jiwoon Park
Julian De Freitas

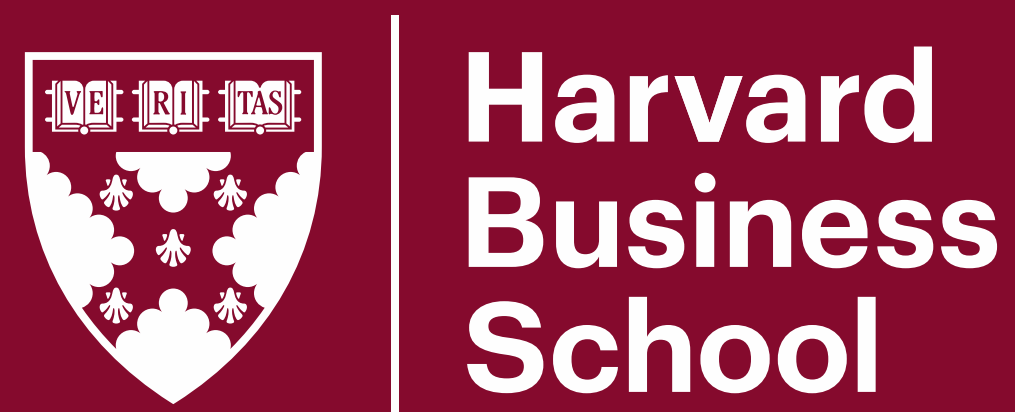

# Automated Comment Moderation Enhances Social Media Advertising Performance

Jiwoon Park
Harvard Business School

Julian De Freitas
Harvard Business School

Funding for this research was provided in part by Harvard Business School.

# Automated Comment Moderation Enhances Social Media Advertising Performance

## Abstract

Social media advertising exposes brands not only to potential customers but also to unfiltered consumer discourse in the form of user comments. While comments can enhance authenticity and engagement, they also introduce reputational risks through spam, hate speech, and negative user-generated content. Despite the increasing prevalence of AI-powered comment moderation solutions, little causal evidence exists on whether moderation (i.e., hiding harmful comments) improves ad effectiveness. Across six empirical studies—including two large-scale field experiments and four online studies—we demonstrate that automated moderation of harmful comments causally improves ad performance, including conversion rates, return on ad spend, and purchase intentions. We also identify two important platform-governance boundary conditions: the gains from moderation depend on whether the platform is transparent about the brand's moderation behavior, and what types of comments are moderated. At the same time, the moderation effect persists when the brand is transparent about its own moderation practices. We advance theory on context effects in social media advertising, by uncovering the first targeted, preventative intervention for avoiding negative adjacencies. For managers, the results show that AI-assisted comment moderation impacts real ad performance but may be contingent upon platform-level transparency design.

The rise of social media has transformed how brands communicate with consumers. Nearly 60% of the world's population uses social media (Kemp 2025), and roughly one third of consumers report they first encounter brands and products through social media (Bedgood 2025). In response, brands and companies invest billions of dollars each year in social media advertising to design campaigns to effectively reach their target audiences and drive purchasing (Suresh and Sharma 2025). However, by design, social media advertisements coexist with commentary on these ads—including spam, misinformation, hate speech, and other harmful discourse that brands neither create nor control. Do such harmful comments hurt advertising performance?

The behavior of both brands and social media platforms would suggest not. Only around 5% of brands spend on dedicated digital solutions to manage comments on their ads (such as by hiding harmful comments), with most of their budget directed to optimizing ads on the social media platforms themselves (De Freitas and Corsi 2025). Furthermore, major platforms such as X and Facebook have scaled back their moderation efforts—often in the name of "free speech" (Duffy 2025) or not stifling innovation (Booth 2025). In contrast, a cohort of Software-as-a-Service (SaaS) platforms has emerged to help companies automatically manage comments on their social media advertising, claiming that doing so improves marketing outcomes. Even so, we are aware of no academic work that empirically tests these claims. While adjacent work in computer science and human-computer interaction has examined platform-level content moderation, such as the efficacy of removing hate speech and toxic content (e.g., Chandrasekharan et al. 2017) or how the users react to their own post removals (e.g., Jhaver et al. 2019), how brand-initiated hiding of harmful ad comments affects consumer perceptions or advertising performance remains an open question.

Here, we draw on both field studies and lab evidence to provide the first causal evidence on whether automated management of harmful ad comments improves ad performance in real social media advertising campaigns. We also test what component of automated comment management—moderation or engagement—drives these effects, and whether they are modulated by transparency about these actions on the part of the social media platform and brand itself.

## Conceptual Framework

### *Contextual Effects in Social Media Advertising*

Work in marketing has sought to identify what makes advertisements effective, finding effects of how the advertising message is framed (e.g., Shiv, Edell and Payne 1997; Wadhwa et al. 2019), the style of the language used to persuade (e.g., Bharti and Sussman 2025; McQuarrie and Mick 1996; Villarroel Ordenes et al. 2019), and how the ad's content is presented (e.g., Hartmann et al. 2021; Jung and Dubois 2023; Wang et al. 2020).

Beyond ad content, an ad's effectiveness depends on its surrounding context (e.g., Bushman 2005; Goldberg and Gorn 1987; Goldfarb and Tucker 2011; Stipp 2018). For example, ads placed in environments cluttered with other ads suffer reduced brand recall (Hammer, Riebe and Kennedy 2009; Nelson-Field, Riebe and Sharp 2013), and various types of ads (banner, native, and video ads) are more likely to be clicked when they are congruent with the surrounding context (Aribarg and Schwartz 2020; Belanche, Flavián and Pérez-Rueda 2017; Goldfarb and Tucker 2011; Shamdasani, Stanaland and Tan 2001). Interpretation of an ad can also be affected by a spillover of affective states from media shown before the ad (Aylesworth and MacKenzie 1998; Goldberg and Gorn 1987). Research on brand safety in digital advertising similarly demonstrates that the content environment surrounding a brand shapes consumer evaluations through contextual spillover: when consumers encounter a brand in environments containing toxic or harmful content, they attribute greater blame to the brand and form less favorable attitudes toward it (Bernritter et al. 2025; Lee, Kim and Lim 2021). This contextual spillover logic motivates our focus on the comment environment beneath a brand's own ads—a context the brand can directly control.

In social media advertising, the context includes both adjacent content in the social media feed and organic comments on the ads themselves. Existing research demonstrates the influence of adjacent content, showing that ads which appear adjacent to negatively valenced content (e.g., a pet ad next to a post raising awareness about animal abuse) elicit negative word of mouth intentions about the brand, because the brand seems to condone or fail to notice this context (Lee, Kim and Lim 2021). These effects also occur when the adjacent content is unrelated (e.g., a clothing ad that appears alongside a news post on the health risks of vaping), because the perceived misalignment between the brand and its context undermines brand trust (Grewal, Stephen and Vana 2025)—the belief that a brand is reliable, acts with integrity, and delivers on its promises and commitments (Chaudhuri and Holbrook 2001; McKnight, Choudhury and Kacmar 2002).

While this past work has provided a foundation to understand the role of context in advertising, it also has several limitations. Firstly, it focuses on adjacencies that are outside of a brand's control, such as posts by unrelated accounts (Grewal, Stephen and Vana 2025; Lee, Kim and Lim 2021), prior media exposure (Belanche, Flavián and Pérez-Rueda 2017; Goldberg and Gorn 1987), or the density of competing advertisements (Nelson-Field, Riebe and Sharp 2013). Even if brands are aware of the unintentional harm caused by these contextual factors, they cannot easily control, correct, or prepare for them in advance. For example, Lee, Kim and Lim (2021) suggest that brand managers demand that platforms help to prevent such adjacencies, yet this is infeasible in practice given that giant platforms like Facebook are unlikely to cater to individual brands, and (if anything) these platforms have been *reducing* moderation on their platforms (Kaplan 2025; McMahon, Kleinman and Subramanian 2025). As another example, Grewal, Stephen and Vana (2025) conclude that brands should monitor their brand media accounts and swiftly apologize for brand safety incidents, yet this is still a reactive solution that does not fully avoid potential brand and financial damage. While the authors do also find that stronger brands are more resilient to context effects, a strong brand can only attenuate these effects but not prevent them from doing some damage altogether.

Second, past work has not measured actual ad performance on real social media websites, but measured attitudes (e.g., brand attitudes, intention to spread negative word of mouth) in reaction to hypothetical incidents, limiting its marketing relevance. While measures of attitudes can be useful proxies of top-of-funnel activity, the gold standard outcome for marketers is still advertising performance, such as website registrations or return on ad spend (ROAS). Furthermore, testing hypothetical incidents alone can artificially make manipulations more salient. As Lee, Kim and Lim (2021) acknowledge, "we programmed the experimental stimuli to appear for a certain amount of time to ensure that participants were exposed to the stimuli without skipping". Only one study, Grewal, Stephen and Vana (2025) looks at real social media data, although their data are descriptive rather than causal and originate from one platform (X).

### ***Automated Comment Moderation and Engagement***

Contrasting with prior work, we focus on the contextual factor of user-generated comments that appear directly under the brand's ads. Unlike other adjacent posts that occur within a user's feed, which vary based on personalized algorithms, an ad's comment section looks the same for all consumers and, crucially, the brand has direct control over these comments: it can engage them ("engagement") and/or hide them ("moderation").

Engagement is when brands respond to certain types of comments, such as FAQs or certain types of negative comments like constructive negative feedback (Saljoughian et al. 2025). Strategic and timely responding to such consumer comments—such as by using appropriate tone and linguistic styles and providing sufficient information—can increase or restore consumers' trust in the brand, by making the brand appear responsive, reliable, and trustworthy (Gao, Rui and Sun 2023; Herhausen et al. 2023; Hill Cummings et al. 2025; Homburg, Ehm and Artz 2015; Tax, Brown and Chandrashekaran 1998).

Moderation is when a brand hides harmful comments, which typically includes either universally harmful comments that are generally viewed as making the thread less pleasant and safe (e.g., profanity, discrimination, or personally identifiable information) or comments that are specifically harmful to the brand in question, such as unconstructive brand attacks (e.g., brand attacks or brand impersonations). Harmful comments do not include constructive negative comments (e.g., product complaints), which are typically treated as engageable comments (De Freitas and Corsi 2025). In this sense, harmful comments on social media also depart from a context like product reviews, where comments are only supposed to reflect relevant product experiences from those who purchased the offering. Likewise, because negative or mixed content in product reviews can carry diagnostic weight and enhance perceived helpfulness or credibility (Rocklage and Fazio 2020; Wu 2013), hiding harmful comments does not deprive consumers of this diagnostic role since product complaints and other constructive negative feedback remain visible and engaged with rather than hidden.

Brands also typically hide harmful comments instead of deleting or engaging with them. Unlike deleted social media comments, hidden harmful comments remain visible to the poster even as they are made invisible to others, decreasing the chance of the poster realizing their comment was hidden. Engaging with harmful comments is also problematic: prior research finds that actively responding to harmful comments like unwarranted insults with tactics like denial or apology fails to build trust and can backfire (Du, Zhou and Cutright 2025). By contrast, prior work on managerial responses to negative reviews finds that responding to such content—rather than ignoring it—improves consumer attitudes and purchase intentions (Esmark Jones et al. 2018; Le and Ha 2021). Hiding harmful comments thus represents a qualitatively different strategy: rather than engaging with criticism, the brand removes it from public view. This distinction also connects to recent work showing that influencers who disable their comment sections entirely are penalized by consumers, who interpret the action as silencing (Daniels and Wu 2024). Our work differs in examining brand-initiated hiding of specific harmful comments—a more targeted and less visible action than wholesale comment disabling—and tests its effects on advertising performance in real field settings.

In sum, past work has not assessed how *proactive, actionable* brand engagement and moderation of harmful ad *comments* impacts financial *ad performance.* This might be because it has only recently become possible, thanks to NLP and generative AI models, to implement such an intervention in an automated manner for an entire ad campaign, 24/7 at scale. We predict:

> **H1**. Turning on automated comment management (versus keeping it off) enhances social media advertising performance.

What aspect of automated comment management most improves ad performance? On the one hand, engagement might chiefly improve ad performance, by showing that the brand is responsive to customer concerns. On the other hand, moderation might chiefly impact brand performance, because harmful comments serve as a strong diagnostic signal of brand negligence or lack of supervision over the advertising environment. To use an analogy, one might still frequent a café where the waiter does not talk much, but not if the windows are broken. This is consistent with prior work in criminology and sociology which argues that visible disorder (e.g., broken windows) signals a lack of control or oversight (Wilson and Kelling 1982). Unmoderated harmful comments might similarly signal a poorly managed brand environment, reducing brand trust (Grewal, Stephen and Vana 2025), in turn hurting ad performance. Thus:

> **H2**. The effect of automated comment management on ad performance is chiefly explained by the effect of comment moderation, not engagement.
>
> **H3**. The effect of comment moderation on advertising performance is explained by perceived brand trust.

### *The Role of Platform and Brand Transparency*

So far, our predictions assume that comment moderation is imperceptible to consumers. Yet, in practice, platforms can make a brand's moderation actions transparent, by allowing consumers to see what comments have been hidden.[1] One possibility is that such platform transparency has no impact on the effect of moderation on ad performance, or the effect is even strengthened, because it leads brands to be lauded for removing harmful comments. A second possibility is that platform transparency eliminates or reverses this moderation effect, because consumers trust the brand less once they learn it is hiding content from them, perhaps because they think the brand is manipulating them into having a falsely positive impression of it (Campbell and Kirmani 2000; Friestad and Wright 1994). This inference would be consistent with the fact that they are learning this information from a third-party, as if the brand itself would otherwise choose to continue hiding its moderation practices from the consumer (Green and Armstrong 2012; Semaan, Kocher and Gould 2018).

[1] As of January 2026, the major social platforms differ in how salient they make hidden or down-ranked comments. X is the most transparent: it offers a clearly labeled "See hidden replies" sub-page, surfaced as a distinct view. TikTok and Instagram offer a similar capability—users can reveal comments that the brand or platform has hidden—but the option is buried; it only appears after scrolling to the very bottom of the comment section. Facebook offers the least transparency. To see all comments, users have to manually switch the comment sort from the default "Most relevant" to "All comments." Even then, Meta's ranking system does not make explicit which comments were hidden.

Whether this negative brand inference is made may also depend on the type of harmful comments that were moderated, which consumers may discover in practice if the platform's transparency feature allows for it (e.g., "See hidden replies"). When the harmful moderated content is *brand-directed* (e.g., hostility aimed at the specific brand), consumers might view moderation as a self-interested brand action, leading moderation to backfire. In contrast, when the moderated content is universally harmful (e.g., hate speech), the presence of a self-interested motive is not as clear, so the moderation effect should not backfire. Thus:

> **H4a.** Platform-initiated transparency of a brand's moderation practices reverses the positive effect of moderation on purchase intentions, if the harmful content was targeted at the brand.
>
> **H4b.** Platform transparency does not reverse the positive effect of moderation on purchase intentions, if the harmful content was universally harmful.

Finally, in practice brands can also transparently disclose their moderation practices themselves, as by posting their policy surrounding such practices at the top of their social media pages. For similar reasons to platform transparency, brand transparency might backfire because consumers learn new, unsavory information about the brand, even if the brand discloses this information itself. On the other hand, brand transparency might not impact the effect of comment moderation on ad performance, since the brand voluntarily reveals this information, thereby counteracting possible perceptions of manipulative intent. Also, in practice brand transparency does not involve showing consumers what comments were hidden, so consumers might not realize that the brand can also hide harmful comments directed at the brand, not just universally harmful comments. Hence:

> **H5.** The positive effect of moderation on purchase intentions is robust to brand-initiated transparency of this practice.

Figure 1 presents the conceptual framework, in which harmful comment moderation influences ad performance through brand trust. This effect is moderated by platform transparency, but only when the type of harmful comment moderated is brand-directed, rather than universally harmful.

**Figure 1.** Conceptual Framework

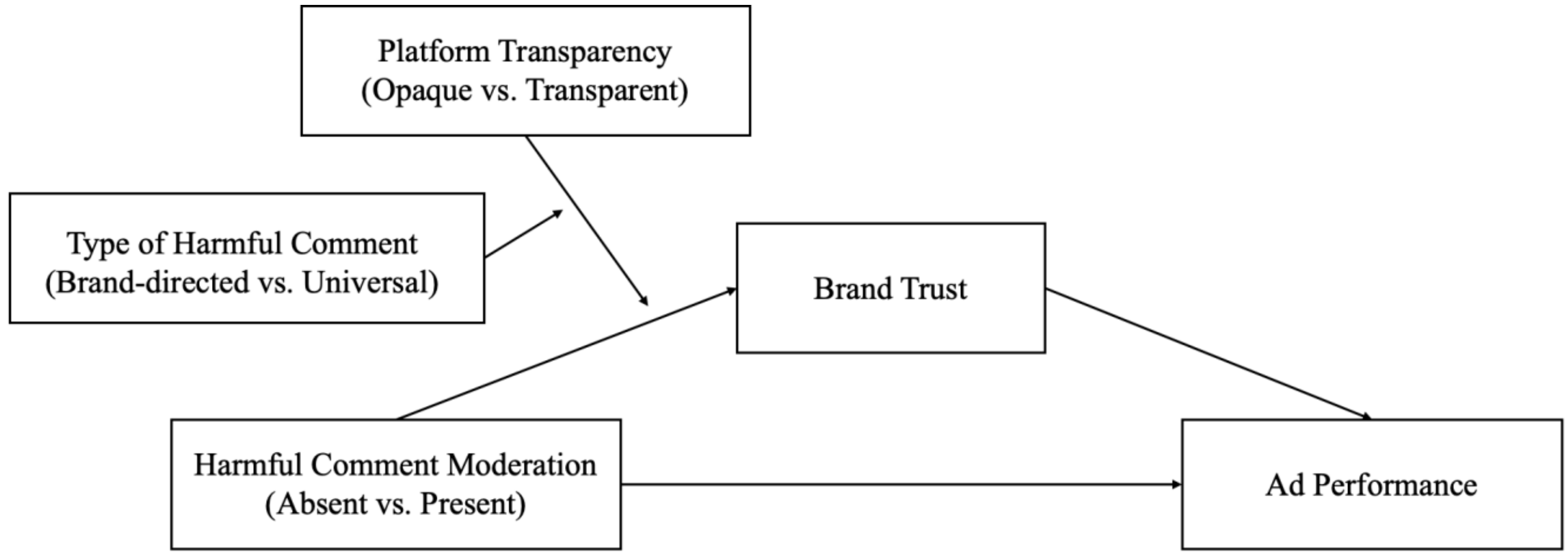

**Overview of Studies**

Two field studies spanning Trustpilot, Instagram, and Facebook (N = 4,374,100), and three experiments with one follow-up replication (N = 4,873) test whether automated comment management of a brand's ads measurably impacts ad performance and purchase intentions. We adopt a two-step strategy. First, Studies 1 and 2 test the combined effect of automated moderation and engagement on ad performance (H1) in the field, reflecting how such interventions are deployed in practice: Study 1 provides motivating evidence using a before-after treatment on Trustpilot, while Study 2 provides a stronger causal test through a concurrent A/B test on Instagram and Facebook. Second, Studies 3–5 then provide evidence from online lab experiments into the specificity, mechanism, moderating conditions, and robustness of these effects. Study 3 tests whether the effect of automated comment management is driven more by comment moderation or engagement. Studies 4 and 5 test the potential moderating impact of platform and brand transparency.

We theoretically add to work on how to avoid negative contextual effects in social media advertising (e.g., Grewal, Stephen and Vana 2025; Lee, Kim and Lim 2021; Stipp 2018), by focusing on ad comments, which are controllable by the brand. We test the *preventative, targeted* intervention of comment management, and determine whether engagement or moderation is chiefly responsible for the effect on ad performance. We also contribute to work on transparency in advertising (Kim, Barasz and John 2019; Semaan, Kocher and Gould 2018; Whittaker et al. 2025), by testing the potential moderating role of platform transparency, and whether it is contingent on the type of harmful comment. We contribute to marketing practice by measuring the role of an actional marketing intervention on real ad performance on real social media platforms, and by testing the role of another considered action in practice—brand transparency.

## Study 1: Large-Scale Comment Moderation and Engagement with Before vs. After Design

Studies 1 and 2 test the combined effect of automated moderation and engagement (H1), using BrandBastion's solution deployed in practice. BrandBastion is an AI-driven SaaS solution for comment management used by global brands. The first field experiment was conducted in partnership with BrandBastion and a financial service company that provides international money transfer and remittance services. After initially deploying its ads without any automated comment management, the financial services company turned on BrandBastion's comment management solution, which consisted of both automated comment moderation and engagement on the brand's behalf (for details, see Web Appendix A). The experiment was conducted over a 10-month period, providing a large-scale behavioral dataset that enables us to examine the impact on ad performance. The primary measure of interest to this brand was ad performance, which was registration completion rate per click, i.e., whether consumers completed a registration form on the brand's website after clicking the ad; impression-level data (e.g., reach) were not available for this study. Note, because the intervention consists of adding automated comment moderation and engagement on the comments—without changing the content, targeting, spend, or other parameters across the treatment and control conditions—this approach allows us to avoid, or at least significantly minimize, the chance of divergent delivery in A/B tests (Burtch et al. 2025). Even so, we also replicate effects in controlled Studies 3–5.

### *Methods*

In October of 2020 the financial services company launched a text-and-image advertisement on Trustpilot, which included a clickable link directing users to the company's website, where consumers could register for an account. In March 2021, five months later, the BrandBastion comment management system was implemented, providing 24/7 moderation and engagement on all comments in the ad's comment section. The moderation protocol involved hiding comments that included harmful content such as spam, scams, violence, inappropriate remarks, or personally identifiable information; all other types of comments remained visible on the brand's social media assets. The engagement protocol involved issuing responses to certain categories of comments that were pre-approved by the brand, including FAQ, purchase intent, customer complaints, and fan community. All other comment types were manually engaged with by the brand, in both conditions.

The control period consists of data from the five months before implementation of the BrandBastion solution (October 11, 2020, to March 10, 2021), and the treatment period consists of data from the five months following it, during which automated management was active (March 11, 2021, to August 1, 2021). Over the course of these ten months, the ad received a total of 3,407,661 link clicks, 62,033 completed registrations, and 15,902 user-generated comments. During the control period, 7,099 comments were submitted in total, and 1,557 company replies were issued. During the treatment period, 8,803 comments were submitted in total, and 1,734 replies were issued. The brand spent $19,993.91 on the campaign in total.

In what follows, we report descriptives, including top-of-funnel behavioral metrics that reflect consumers' direct responses to the ad, followed by bottom-of-funnel performance metrics: including number of link clicks, number of web registrations completed, and registration completion rate per link click. The last variable served as our primary outcome variable. Due to how data were provided by Trustpilot, we provide only aggregated metrics for each period.

***Results***

Table 1 presents descriptives for the control and treatment groups for the over 1 million interactions spanning the extended study period. As a manipulation check, we first examined whether automated moderation altered the sentiment of visible comments as intended. Because raw comment data were not available for Study 1, we used sentiment measures that were generated by BrandBastion's AI-based classification system, which reflect the aggregated distribution of comment sentiment that appeared under the ads during the control and treatment periods. The categories of comments treated as harmful were co-developed between BrandBastion and the brand and remained constant; the sentiment difference therefore reflects BrandBastion's removal of harmful content, not differential definitions of what counted as harmful. We observed a significant reduction in the proportion of negative sentiment comments visible during the treatment period (from 57.37% to 43.06%; $X^2(1) = 222.67$, $p < .001$). Conversely, positive sentiment among visible comments significantly increased (from 8.98% to 19.36%; $X^2(1) = 251.5$, $p < .001$). We note that this shift in positive and negative comments could reflect not just the direct effect of hiding harmful comments, but also how this intervention influenced new comments from subsequent users, although we are agnostic given that we focus on how the treatment affects ad performance.

**Table 1.** Descriptive data for Study 1.

| | Control Period (5 months) | Treatment Period (5 months) |
|---|---:|---:|
| **User- and Brand-Generated Comments** | | |
| # Comments Monitored (n) | 7,099 | 8,803 |
| # Harmful Comments Removed (n) | 730 | 4,051 |
| Negative Sentiment (%) | 57.4 | 43.1 |
| Positive Sentiment (%) | 8.98 | 19.36 |
| # Replies Issued (n) | 1,557 | 1,734 |
| **Ad Performance** | | |
| # Link Clicks (n) | 1,522,246 | 1,885,415 |
| # Website Registrations Completed (n) | 25,469 | 36,564 |
| Registration Completion Rate (%) | 1.67 | 1.94 |

Next, we examined the impact of the automated comment management intervention on ad performance. Descriptive analyses showed an increase in the total number of link clicks in the treatment period (1,885,415 clicks) compared to the control period (1,522,246 clicks; a 23.9% increase). Similarly, the number of website registrations completed was higher during the treatment period (36,564 completions) than the control period (25,469 completions; a 43.6% increase). Most importantly, registration completion rate per link clicked significantly increased for the treatment period (1.94%) relative to the control period (1.67%; $X^2(1) = 333.76$, $p < .001$), showing that the treatment condition improved the main target performance metric.

***Discussion***

Study 1 provides evidence that automated comment management (consisting of both comment moderation and engagement) improves the performance of a social media ad campaign on Trust Pilot, increasing a user's likelihood of completing a web registration upon clicking on the ad link. Importantly, we found this effect of comment management while controlling for all other aspects of the campaign itself, except time itself (given the before-versus-after comparison). Additionally, given the sequential nature of the design, consumers in the treatment period may have had prior exposure to the brand's ads during the control period. Study 2 tests whether the results can be conceptually replicated using an experimental (i.e., A/B) design that controls for time, with a different brand on different social media platforms.

## Study 2: Generalization Test with A/B Design

The second field experiment was conducted in partnership with BrandBastion and an online personal growth platform that offers various courses, programs, and content to help individuals enhance their lives. BrandBastion collaborated with this company to conduct a controlled A/B test of its solution: the experiment compared the performance of two pairs of ad campaigns—one with and one without comment management (i.e., the combined moderation and engagement protocols), which ran on Facebook and Instagram. Study 2 offers three key advantages.

First, the use of two social media platforms tests generalization beyond a single platform. Second, the A/B testing framework removes the potential temporal confound present in Study 1. Third, this time the primary outcome metric we study is return on ad spend (ROAS), a direct monetary measure of performance beyond the click-to-registration rates in Study 1. Thus, the campaign is particularly relevant to marketing practitioners, who in practice must financially justify investment of marketing spend to other stakeholders in the business, such as the Chief Financial Officer.

As in Study 1, we report descriptives for the dataset, and the primary performance metric that was targeted (in this case, ROAS). Because for this study additional variables were also collected—including user engagement, number of likes, comments, and shares of the post, as well as the user-generated comments—we also report these for completion.

***Methods***

Two pairs of campaigns were created, for the treatment and control groups on each platform (Instagram and Facebook). The campaigns were identical in targeting, creative, setup, budget, and ad expenditure, and ran concurrently from April 4, 2023 to May 2, 2023 on both platforms. The conditions differed only in whether BrandBastion's comment management system was active for ad comments: in the control group, comments were not managed, whereas in the treatment group, the comment moderation solution hid certain comments (e.g., spam, scams, hate speech, brand attacks and critiques, and violent or inappropriate comments) and the engagement system automatically responded to certain comments (e.g., fan community posts, frequently asked questions, feedback and complaints) based on predefined engagement matrices and brand-provided response guidelines.

***Results***

As there were no significant differences between campaigns run on Instagram and Facebook, we report results based on aggregated data across platforms. Table 2 presents descriptives for the control and treatment groups. The similar levels of impressions and reach across conditions suggest that algorithmic delivery was not meaningfully affected by the comment management intervention—which is consistent with our design holding ad content and targeting constant (unlike most A/B designs; Burtch et al. 2025).

As a manipulation check, we again examined how the comment environment changed after implementing the automated comment management system. We classified raw comments as positive, negative, or neutral using Hugging Face's RoBERTa model. The manipulation check involved two steps. First, we confirmed that the treatment group had significantly fewer visible comments than the control group ($b$ = -1.10, SE = 0.37, $z$ = -3.01, $p$ = .003). Second, we verified that the moderation system was specifically targeting negative comments for hiding, rather than comments of other valences, by estimating Firth-penalized logistic regression models predicting whether a given comment was hidden: negative comments in the treatment group were significantly more likely to be hidden than those in the control group ($b$ = 5.44, 95% CI = [2.24, 11.23]). In contrast, positive comments did not significantly differ in hiding rates between groups ($b$ = 1.67, 95% CI = [-1.40, 6.71]). We also observed a small but statistically significant difference in engagement rates per ad reach (treatment: 74.7%, control: 74.2%; $\chi^2(1) = 39.25$, $p < .001$).

**Table 2.** Descriptive results for Study 2

| | **Control Group** | **Treatment Group** |
|---|---:|---:|
| **User Engagement** | | |
| Impressions (n) | 545,627 | 529,774 |
| Post Comments (n) | 31 | 16 |
| Post Engagement (n) | 362,391 | 357,094 |
| Engagement Rate (%) | 74.2 | 74.7 |
| Post Reactions (n) | 1,137 | 1,369 |
| Post Saves (n) | 160 | 197 |
| Post Shares (n) | 135 | 159 |
| Website Content Views (n) | 31 | 72 |
| Searches (n) | 27 | 63 |
| **Ad Performance** | | |
| Reach (n) | 488,581 | 477,858 |
| Clicks (n) | 18,463 | 17,791 |
| Amount Spent (USD) | 9,998.44 | 9,995.47 |
| Adds to Cart (n) | 40 | 44 |
| Purchases (n) | 15 | 22 |
| Purchase Conversion Value (USD) | 4,585 | 6,778 |
| Return on Ad Spend (ROAS) | 0.459 | 0.678 |

Notes: Impressions refers to the total number of times the advertisement was viewed by users, while reach refers to the number of unique users who were exposed to the advertisement. Post Reaction refers to the number of emoji-based reactions to the ad. Note that the fact ROAS < 1 suggests the campaign lost money overall, even though the comment moderation solution improved its performance.

We report the primary ad performance metric of return on ad spend (ROAS)—calculated as purchase conversions per amount spent on the ad (USD)—descriptively because ROAS reduces to a single ratio per condition. Consistent with our predictions, ROAS was descriptively higher in the treatment group (0.678) than the control group (0.459). The treatment group generated more purchases (22 vs. 15) at a lower cost per purchase ($454.34 vs. $666.56), reflecting a meaningful financial return for the same ad spend.

***Discussion***

By using a more controlled A/B experimental design on two new social media platforms, we again find evidence that automated comment management improves social media ad performance: compared to the control group, ad campaigns with automated moderation and engagement generated substantially greater monetary return per dollar spent. This study shows generalization to new social media platforms, and strengthens the internal validity of our findings, by providing evidence from a concurrent A/B design in which automated comment management is the only difference between conditions. Together, results from two large-scale field experiments show that automated comment management (moderation and engagement) improves social media advertising performance.

## Study 3: Does Moderation or Engagement Improve Ad Performance More?

While Studies 1-2 provide causal evidence that a solution consisting of both automated moderation and engagement improves ad performance, it does not isolate how much each of these comment management tactics contributes to ad performance. In practice, brands implement both simultaneously (as in Studies 1 and 2), under the assumption that both improve ad performance, but is this assumption correct? Since in practice brands are not interested in testing this, we answer this question by utilizing controlled lab experiments. Consistent with prior experimental studies of advertising effectiveness (Chan and Ilicic 2019; Ostinelli and Luna 2022; Paharia 2020), we measure purchase intentions as a proxy of ad performance.

***Methods***

This study was pre-registered (https://aspredicted.org/59z8-mtgm.pdf). Since the stimulus consisted of a fictitious sponsored post for a hair dryer, we recruited 400 female participants from the US via Cloud Research who passed attention checks. Participants who failed comprehension checks were excluded from analyses, leaving a final sample of 398 participants (97.7% self-identified females, $M_{age}$ = 40.5 years old). Although nine participants did not self-identify as female in the survey (3 male, 3 other, and 3 prefer not to disclose), we retained them in the analyses as the recruitment was targeted at females on the Cloud Research platform, meaning only females should have seen it. Excluding these participants did not change the results.

Study 3 employed a 2 (brand engagement: present vs. absent) × 2 (harmful comment moderation: present vs. absent) between-subjects design. All participants viewed a sponsored post for a hair dryer from a fictitious brand named Aetherflow, which included a "Shop now" button. The post was accompanied by a set of user-generated comments comprising two positive comments, reflecting fan community style messages that express affinity toward the product, and two harmful comments, reflecting brand attacks that contain profanity without offering constructive feedback or meaningful questions. Which of these comments participants saw depended on condition. In the Brand Engagement Present condition, the brand visibly replied to the positive user-generated comments. In the Comment Moderation Present condition, harmful comments were not visible. In the respective absent conditions, the brand did not respond to comments, and harmful comments remained visible. For realism, the user interface of the post and comment sections was modeled after Instagram. See Figure 2 for all four stimuli.

After viewing the social media advertisement and the associated comments, participants were asked to imagine that they were shopping for a new hair dryer and to indicate how likely they would be to click on the "Shop now" button on a 100-point scale (1= Not at all likely, 100 = Extremely likely). On the next page, participants completed a comprehension check which asked them to correctly identify the product advertised on the previous page.

**Figure 2.** Stimuli Used in Study 3

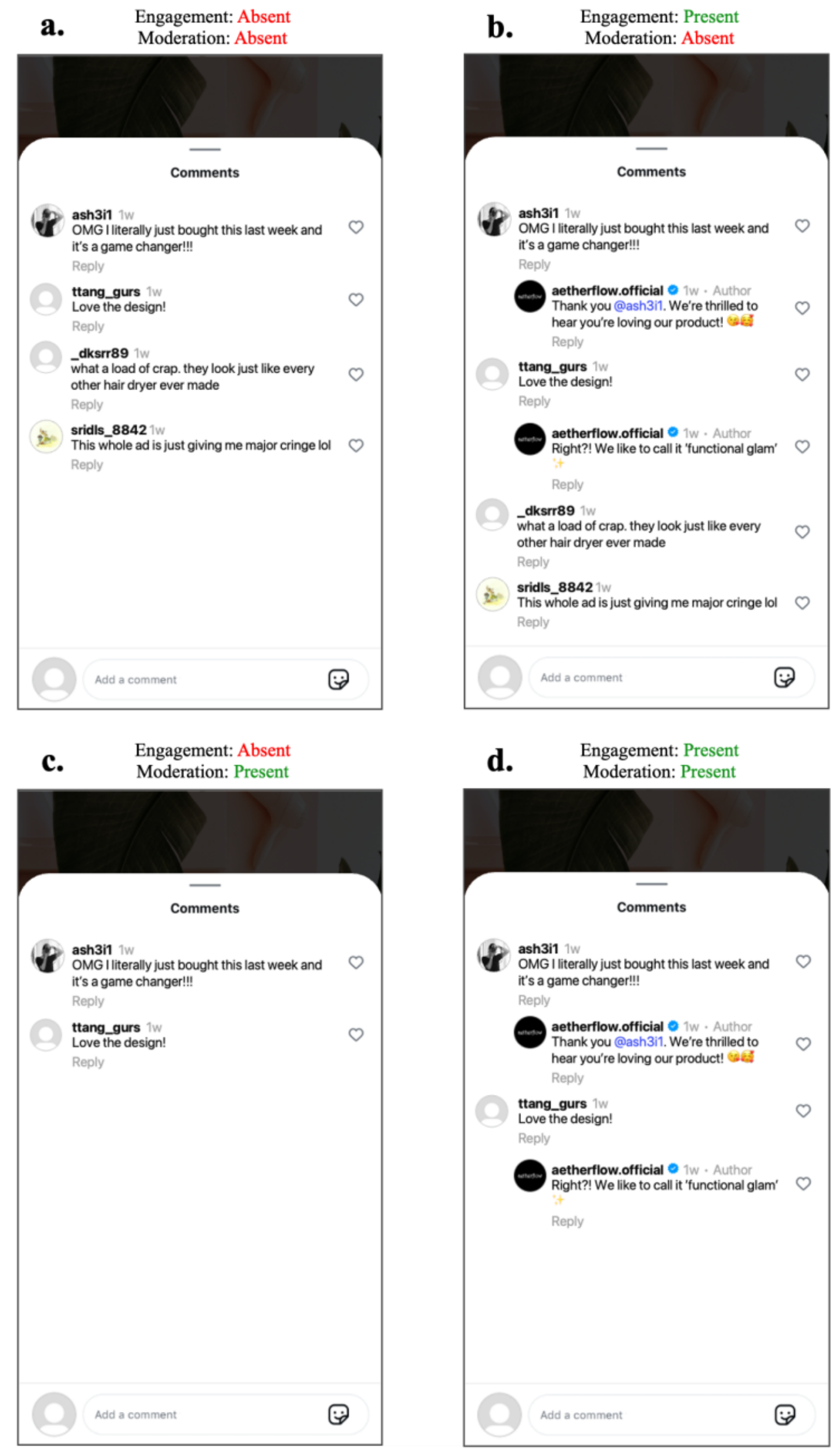

## ***Results***

We found a main effect of comment moderation ($F(1, 394) = 14.14$, $p < .001$, $\eta^2 = 0.03$), but no main effect of brand engagement ($F(1, 394) = 1.56$, $p = .213$, $\eta^2 = 0.004$) nor an interaction effect ($F(1, 394) = 1.95$, $p = .164$, $\eta^2 = 0.005$). Participants said they were more likely to click on the "Shop now" button when moderation was present ($M = 47.2$, $SD = 30.3$) than absent ($M = 36.9$, $SD = 28.2$)—Figure 3.

**Figure 3.** Results of Study 3

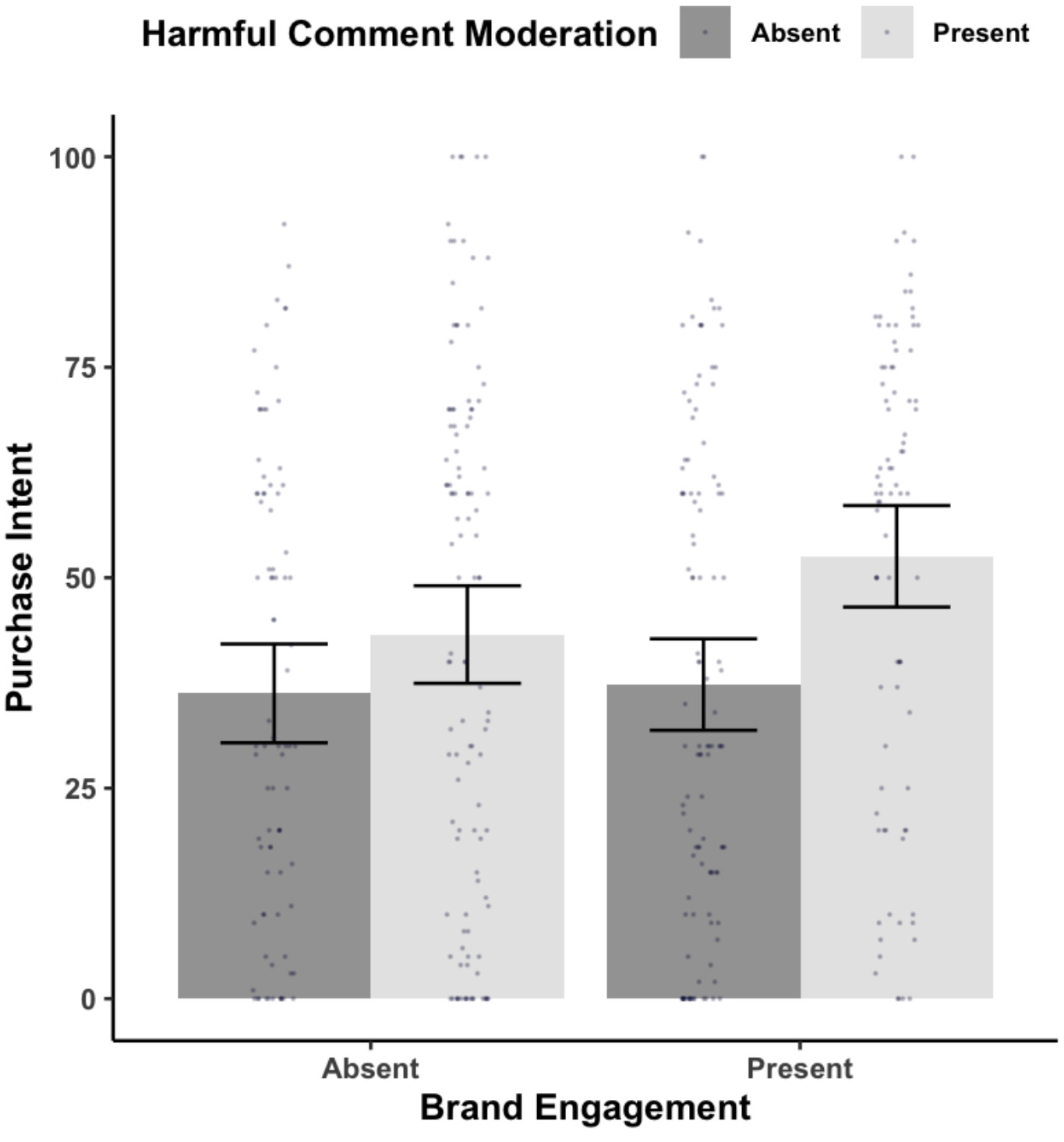

Notes: * $p < .05$, ** $p < .01$, *** $p < .001$

One alternative explanation for these results is that conditions differ in the volume of comments displayed and therefore in information load (Scammon 1977; Wilkie 1974), which could in principle drive differences in purchase intent independent of the comment content (e.g., Iyengar and Lepper 2000. However, the pattern of Study 3 results is inconsistent with this account. The conditions with equivalent comment volume—Engagement Absent / Moderation Absent (Figure 2a) and Engagement Present / Moderation Present (Figure 2d)—both contain four comments, yet differ substantially in purchase intent ($M = 36.3$ vs. $M = 52.5$; $t(168) = -3.85$, $p < .001$, $d = 0.59$). Likewise, when comparing the condition with the fewest visible comments (i.e., Engagement Absent / Moderation Present; Figure 2c) to the condition with the most visible comments (i.e., Engagement Present / Moderation Absent; Figure 2b), purchase intent did not differ significantly ($M$ = 43.3 vs. $M$ = 37.3; $t$(224) = 1.48 $p$ = .140 $d$ = 0.20). These patterns suggest the effect is driven by *what* is shown rather than by *how much* is shown.

***Discussion***

Study 3 suggests that the effect of social media management on social media advertising performance works primarily through the effect of moderating harmful comments, in that only hiding such comments increased purchase intentions.

While our pre-registration indicated that we were agnostic to whether we would find an interaction effect, we did wonder whether we were simply underpowered to detect one, so we conducted a post hoc power analysis to estimate the sample size needed to detect this effect if it exists. Based on the observed effect size in Study 3 (Cohen's $f^2$ = 0.005), the power analysis revealed that a minimum of 399 participants per condition would be required to detect the interaction effect with 80% power at a significance level of $\alpha$ = 0.05. In a conceptual replication of Study 3, we very conservatively employed a sample six-times larger than Study 3: 2,485 US-based female participants from Cloud Research (96.7% self-identified females, $M_{age}$ = 40.6 years old). Once again, we found a main effect of harmful comment moderation $F(1, 2481) = 126.95$, $p < .001$, $\eta^2 = 0.05$), but no main effect of brand engagement ($F(1, 2481) = 1.85$, $p = .174$, $\eta^2 < 0.001$) nor an interaction effect ($F(1, 2481) = 0.20$, $p = .652$, $\eta^2 < 0.001$). Participants were more likely to click on the "Shop now" button when moderation was present ($M$ = 46.9, $SD$ = 29.4) than absent ($M$ = 34.1, $SD$ = 27.2).

While we did not find an effect of brand engagement in this study, it remains possible that such an effect is present in an actual social media setting; even if so, our results suggest that comment moderation exerts the larger effect. Further, even if brand engagement does not improve ad effectiveness, it might yield other benefits beyond ad performance, such as improving the social value of consumers on the brand's own social media pages (e.g., via social word of mouth). Study 4 focuses on understanding how and when comment moderation increases purchase intentions.

## Study 4A: Moderating Role of Platform Transparency for Brand Attack Comments

Studies 4A and 4B serve two purposes. First, following prior work on negative adjacencies, it tests brand trust as the mechanism through which comment moderation influences ad performance (H3). Second, it tests whether and when platform transparency weakens this effect, in the cases where the hidden harmful comments are directed at the brand (H4a, Study 4a) and universally harmful (H4b, Study 4b). Third, we also measure platform trust in addition to brand trust, given platforms are presumably transparent in the first place because they assume this practice increases consumer trust in the platform; we test whether this is in fact the case.

***Methods***

Study 4A was pre-registered (https://aspredicted.org/vxfb-gnhc.pdf). We recruited 802 female U.S. participants who passed the attention checks via Cloud Research and removed those who failed the comprehension check, leaving a total of 800 female U.S. participants remained for analysis (95.6% self-identified females, $M_{\text{age}}$ = 37.9 years old). The study employed a 2 (harmful comment moderation: absent vs. present) × 2 (platform transparency: opaque vs. transparent) between-subjects design, with harmful comments directed at the brand. The manipulation of harmful comment moderation followed the same procedure as Study 3, including the use of harmful comments that were brand-directed. We additionally implemented the platform transparency manipulation by introducing a "See hidden replies" feature. In the Platform Transparent condition, participants were explicitly informed by the platform whether comments had been hidden. When comment moderation was absent, the platform also displayed the message, "This post doesn't have any hidden replies" and the "Hidden replies" page was empty. When moderation was present, the platform also displayed the message, "Some replies were hidden by the author" and the hidden replies were listed; participants did not need to click to see this information—Figure 4. It was explicitly stated that the ad post was authored by the brand Aetherflow and that any replies on the "Hidden replies" page were hidden by the brand. This user interface was modeled after X to increase realism, making it clear that the platform was supplying this transparency information. In the Platform Opaque condition, no information about comment moderation was provided, so participants were unaware if comments had been hidden.

Finally, participants rated their intentions to click on the ad to purchase the product, on a 100-point scale. They then rated their perceived brand trust using three items adapted from Chaudhuri and Holbrook (2001): "I trust this brand", "I would rely on this brand", "This brand is safe" (0 = strongly disagree, 100 = strongly agree; $\alpha$ = .92). On the following page, they rated platform trust, using the same items adapted to refer to the platform instead ($\alpha$ = .94).

**Figure 4.** Stimuli for Comment Sections Used in Study 4A

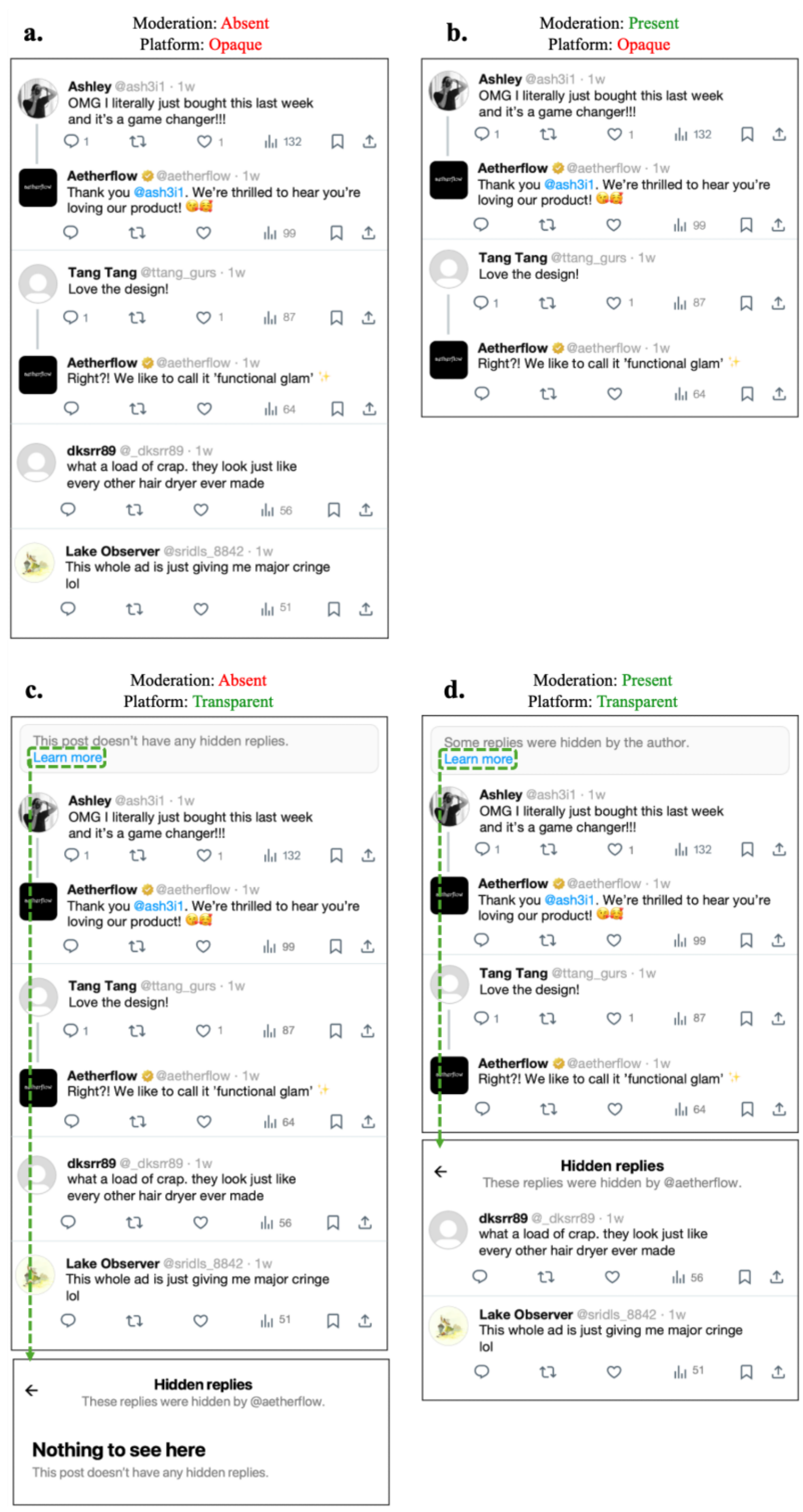

Notes: The "Hidden replies" page in the platform transparent conditions was displayed on the same page as the comments, but further down on the page. The dotted lines in the figure, added here for illustrative purposes only (not present in the original), indicate the relationship between the comment section and the corresponding "Hidden replies" section. Note that the comment section always includes engagement to ensure ecological validity (even though Study 3 demonstrated that brand engagement had minimal impact on purchase intentions), since otherwise the comment

section would appear either blank or consist of only harmful comments (depending on moderation condition).

***Results***

A 2×2 ANOVA revealed no main effect of comment moderation ($F(1, 796) = 1.16$, $p = .282$, $\eta^2 = 0.001$), a main effect of platform transparency ($F(1, 796) = 6.85$, $p = .009$, $\eta^2 = 0.009$), and a significant interaction effect ($F(1, 796) = 24.78$, $p < .001$, $\eta^2 = 0.03$). Follow-up pairwise comparisons showed that when the platform was opaque, purchase intentions were higher when comment moderation was present ($M = 52.6$, $SD = 28.8$) than absent ($M = 40.6$, $SD = 27.5$ $t(796) = -4.29$, $p <.001$, $d = 0.42$), replicating previous results. When the platform was transparent, this effect reversed: purchase intentions were *lower* when comment moderation was present ($M = 37.5$, $SD = 27.0$) than absent ($M = 45.2$, $SD = 28.3$; $t(796) = 2.75$, $p = .006$, $d = 0.28$). Consistent with H4a, platform transparency reversed the positive effect of moderation of harmful comment that was brand-directed—see Figure 5.

**Figure 5.** Purchase intention results for Study 4A

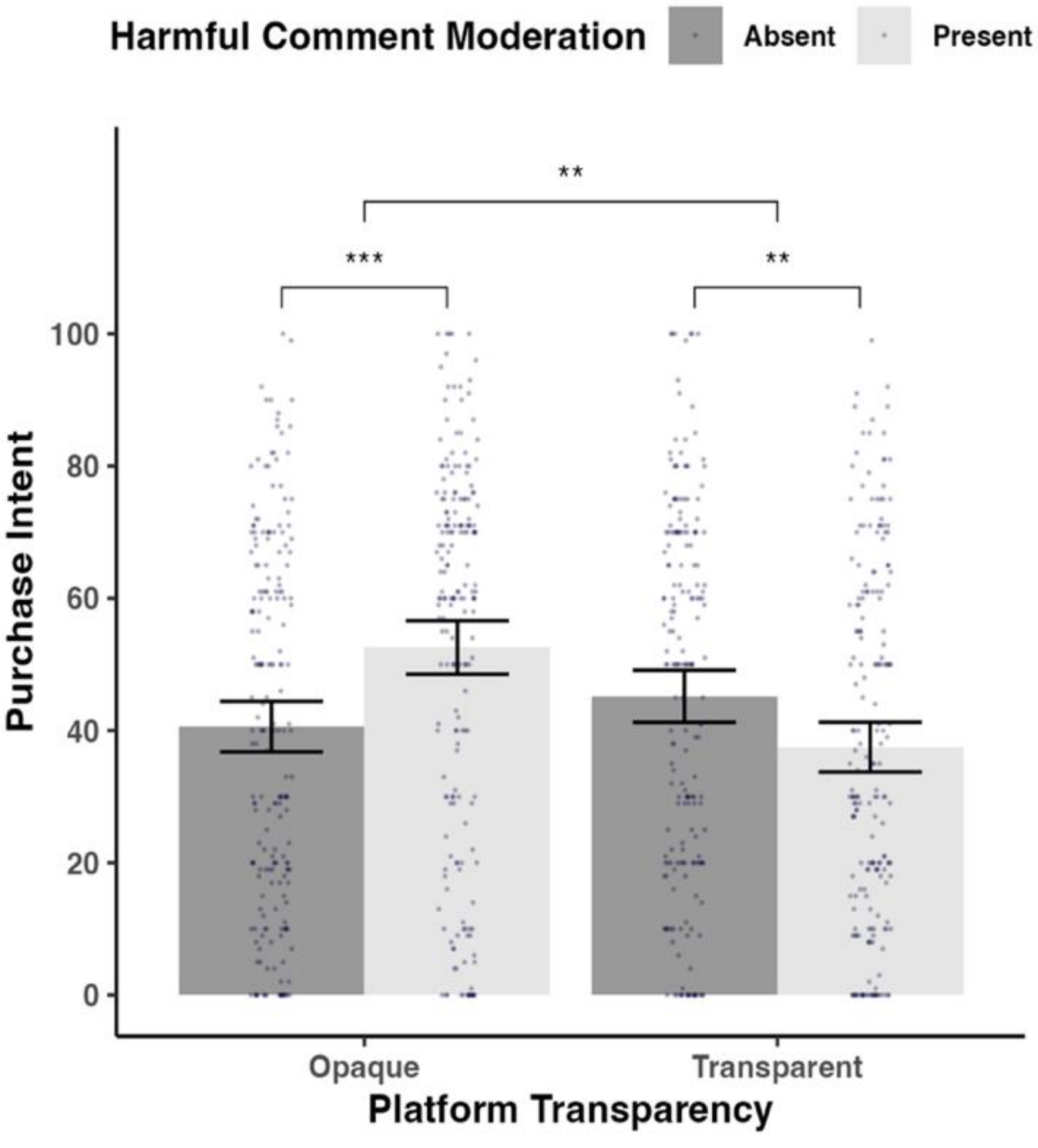

Notes: * $p < .05$, ** $p < .01$, *** $p < .001$

*Perceptions of Brand and Platform Trust*

Results for brand trust mirrored purchase intent. A 2×2 ANOVA revealed no main effect of comment moderation ($F(1, 796) = 0.66$, $p = .416$, $\eta^2 < 0.001$), a marginally significant main effect of platform transparency ($F(1, 796) = 2.97$, $p = .085$, $\eta^2 = 0.004$), and a significant interaction effect ($F(1, 796) = 46.27$, $p < .001$, $\eta^2 = 0.05$). When the platform was opaque, brand trust was higher when comment moderation was present ($M = 53.7$, $SD = 21.3$) than absent ($M = 44.9$, $SD = 20.2$; $t(796) = -4.24$, $p < .001$, $d = 0.43$). When the platform was transparent, participants reported *lower* perceptions of brand trust when comment moderation was present ($M = 41.1$, $SD = 21.1$) than absent ($M = 52.3$, $SD = 20.6$; $t(796) = 5.38$, $p < .001$, $d = 0.54$).

Surprisingly, platform trust showed a similar interaction pattern. A 2×2 ANOVA revealed no main effect of comment moderation ($F(1, 796) = 2.29$, $p = .131$, $\eta^2 = 0.003$), no main effect of platform transparency ($F(1, 796) = 1.72$, $p = .190$, $\eta^2 = 0.002$), and a significant interaction effect ($F(1, 796) = 23.69$, $p < .001$, $\eta^2 = 0.03$). When the platform was opaque, platform trust was higher when comment moderation was present ($M = 50.9$, $SD = 23.6$) than absent ($M = 45.5$, $SD = 21.6$; $t(796) = -2.38$, $p = .018$, $d = 0.24$). When the platform was transparent, platform trust was *lower* when comment moderation was present ($M = 40.9$, $SD = 22.4$) than absent ($M = 51.2$, $SD = 24.0$; $t(796) = 4.51$, $p < .001$, $d = 0.45$). In short, platform transparency backfires for both brand and platform trust.

*Moderated Mediation Analysis*

We conducted a moderated mediation analysis using PROCESS Model 7 (Hayes 2017) with 10,000 bootstrapped samples, with platform transparency moderating the 'a path' from comment moderation to the two competing parallel mediators. The results revealed a significant index of moderated mediation for brand trust (index of moderated mediation = -19.00, 95% BootCI = [-24.71, -13.43]). When the platform was opaque, harmful comment moderation increased click intentions through higher brand trust (effect = 8.37, 95% BootCI = [4.51, 12.26]). In contrast, when the platform was transparent, harmful comment moderation reduced brand trust, leading to lower click intentions (effect = -10.63, 95% BootCI = [-14.69, -6.73]). For platform trust, however, the moderated mediation index was not significant (index of moderated mediation = -1.04, 95% Boot CI = [-2.55, 0.20])—see Figure 6. For robustness, we re-estimated the model using PROCESS Model 8. The results were substantively identical, with moderated mediation through brand trust remaining significant (index = -18.95, 95% BootCI = [-24.72, -13.39]) while the moderated mediation through platform trust again did not reach significance (index = -1.03, 95% BootCI = [-2.54, 0.20]). The added interaction term on the direct path was not significant ($b = 0.33$, $p = .903$).

**Figure 6.** Moderated Mediation Results for Study 4A

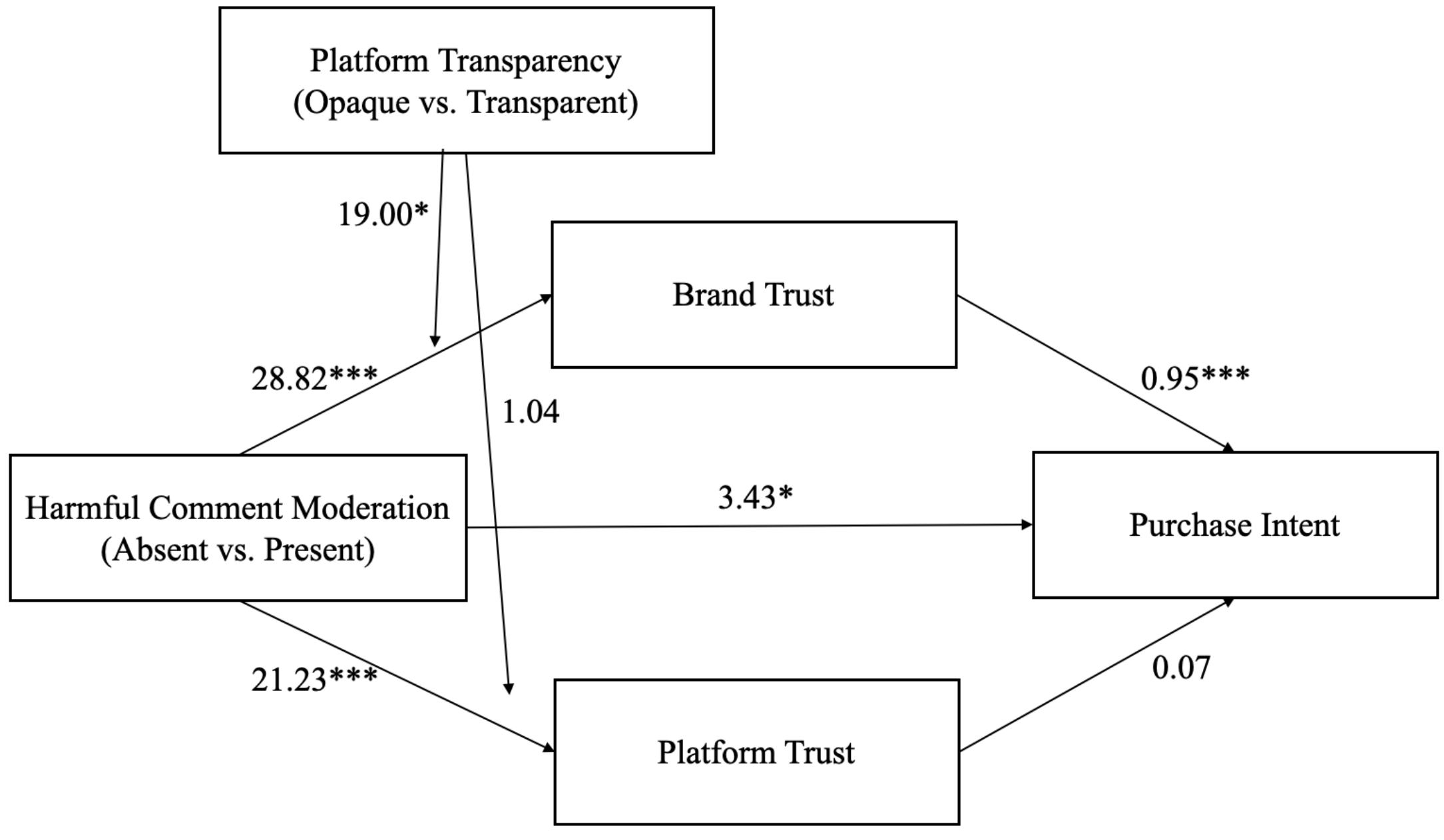

Note: * $p < .05$, ** $p < .01$, *** $p < .001$

***Discussion***

Study 4A yields several useful findings. When the moderated harmful comments were directed at the brand, platform-initiated transparency reversed the effect of comment moderation on purchase intentions—consistent with a backfiring effect (H4a). This moderating effect selectively operated on brand trust, rather than platform trust, which makes sense given that it is ultimately the brand that consumers would purchase from (H3).

Surprisingly, we also found that platform transparency reduces consumers' trust in the platform itself. Even though platform transparency is in consumers' interests, perhaps they penalize the platform for enabling brands to manipulate their impressions of the brand in the first place. Although this inference did not translate into reduced purchase intentions, it suggests platform transparency has counterintuitive consequences for platforms too.

**Study 4B: Boundary On Platform Transparency of Universally Harmful Comments**

Study 4B tests whether the moderating role of platform transparency applies when the harmful comment is universally harmful (e.g., profanity, hostility) rather than brand-directed. Our pre-registration (https://aspredicted.org/s5733h.pdf) predicted that platform transparency would attenuate the moderation effect; we refine this prediction here by distinguishing between comment types, predicting that no interaction effect will emerge for universally harmful comments, since removing such comments is less clearly a self-interested brand action (H4b).

***Methods***

We recruited 1,178 female U.S participants who passed the attention checks via Prolific. After excluding participants who failed the comprehension check, the final sample consisted of 1,173 participants (99.4% self-identified females, $M_{age}$ = 43.4 years old). The design and procedure were identical to Study 4A, except the harmful comments contained explicit language unrelated to the brand or its advertisement—see Figure 7. As in Study 4A, participants viewed the ad post and the corresponding replies pages, then rated their purchase intentions, brand trust ($\alpha$ = .92), and platform trust ($\alpha$ = .93) on 100-point scales.

**Figure 7.** Stimuli for Comment Sections Used in Study 4B

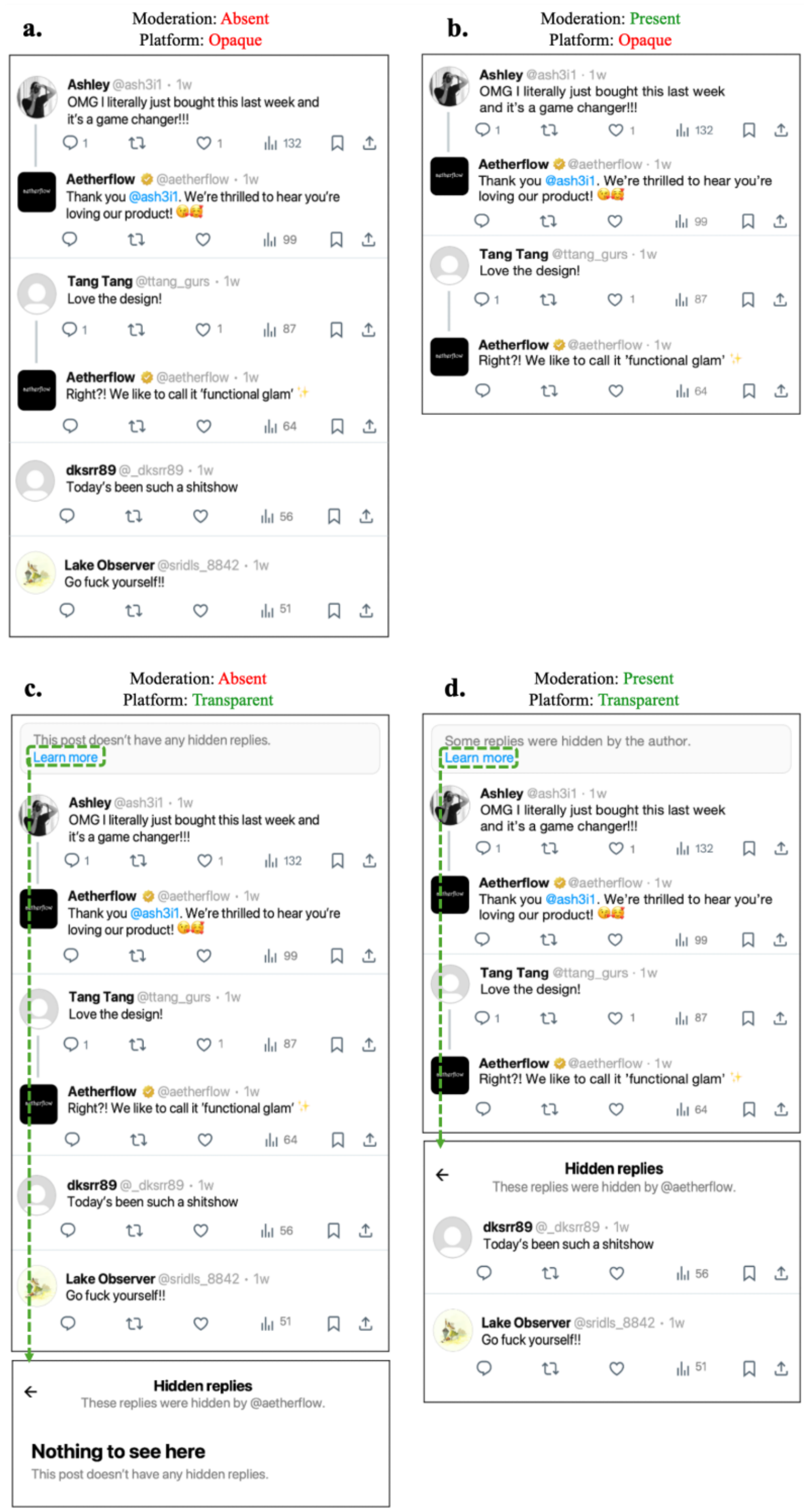

### *Results*

This time, a 2×2 ANOVA on purchase intentions revealed a significant main effect of comment moderation ($F(1, 1169) = 4.87$, $p = .028$, $\eta^2 = 0.004$), no main effect of platform transparency ($F(1, 1169) = 1.61$, $p = 0.205$, $\eta^2 = 0.001$), and no interaction effect ($F(1, 1169) = 1.65$, $p = .199$, $\eta^2 = 0.001$). Regardless of platform transparency, purchase intentions were higher when comment moderation was present ($M = 53.3$, $SD = 29.1$) than absent ($M = 49.6$, $SD = 28.1$). Consistent with H4b, the positive effect of moderation was preserved even under platform transparency when the moderated comments were universally harmful.

Although the interaction was not significant, we conducted pairwise comparisons as preregistered. We found that when the platform was opaque, purchase intentions were higher when moderation was present ($M = 55.4$, $SD = 28.9$) than absent ($M = 49.6$, $SD = 28.2$; $t(1169) = -2.47$, $p = .014$, $d = 0.20$). When the platform was transparent, however, the effect of moderation was no longer significant: purchase intentions were similar when moderation was present ($M = 51.2$, $SD = 29.1$) and absent ($M = 49.6$, $SD = 27.9$; $t(1169) = -0.65$, $p = .513$, $d = 0.05$). Thus, although the overall benefit of moderation was preserved rather than reversed (H4b), it was still descriptively attenuated under platform transparency.

*Perceptions of Brand and Platform Trust*

Similarly, a 2×2 ANOVA on brand trust revealed a main effect of comment moderation ($F(1, 1169) = 3.88$, $p = .049$, $\eta^2 = 0.003$), no main effect of platform transparency ($F(1, 1169) = 0.135$, $p = .713$, $\eta^2 = 0.0001$), and no interaction effect ($F(1, 1169) = 1.178$, $p = .278$, $\eta^2 = 0.001$). Participants reported higher perceptions of brand trust when comment moderation was present ($M = 55.8$, $SD = 20.8$) than when absent ($M = 53.5$, $SD = 20.2$).

Likewise, platform trust showed a main effect of comment moderation ($F(1, 1169) = 10.36$, $p = .001$, $\eta^2 = 0.009$), no main effect of platform transparency ($F(1, 1169) = 0.073$, $p = .787$, $\eta^2 < 0.001$), and no interaction effect ($F(1, 1169) = 0.146$, $p = .703$, $\eta^2 < 0.001$). Platform trust was higher when moderation was present ($M = 50.3$, $SD = 23.2$) than absent ($M = 45.9$, $SD = 23.0$).

*Mediation Analysis*

As we did not find a significant interaction between comment moderation and platform transparency, we conducted a simple mediation analysis using PROCESS model 4 (Hayes 2017) with 10,000 bootstrapped samples, with brand trust and platform trust as parallel mediators. Both brand trust (effect = 2.16, 95% BootCI = [0.02, 4.28]) and platform trust (effect = 0.75, 95% BootCI = [0.26, 1.40]) significantly mediated the effect of comment moderation on purchase intentions. However, brand trust exhibited a substantially stronger association with purchase intentions ($b = 0.92$, 95% CI = [0.84, 0.99], standardized coefficient = 0.66) than platform trust ($b = 0.17$, 95% CI = [0.11, 0.24], standardized coefficient = 0.14), indicating that brand trust is the more consequential pathway from moderation to consumers' behavioral intentions.

***Discussion***

Taken together, Studies 4A and 4B clarify when platform transparency undermines comment moderation. Platform transparency only significantly reversed the impact of comment moderation on purchase intentions when the hidden comments were directed at the brand (Study 4A), rather than universally harmful (Study 4B). At the same time, the moderation effect was relatively flattened even when there was transparency of universally harmful comments, suggesting brands are still best off when the platform is not transparent at all.

## Study 5: The Effect of Brand Transparency

As noted earlier, in practice brands also have the option of being transparent about their moderation practices. While this activity does not afford a theoretically controlled comparison to platform transparency—since in practice brands never reveal exactly what comments were hidden, nor would it be feasible to do so on a continual basis—for practical reasons marketers are nonetheless interested in whether brand transparency impacts ad performance. We predict that brand transparency does not modulate the impact of moderation on purchase intentions, likely because it seems less manipulative (coming from the brand) and because consumers might be unlikely to realize the brand can also moderate brand attacks.

***Methods***

This study was pre-registered (https://aspredicted.org/j5cy4i.pdf). We recruited 1,184 female U.S. participants via Prolific who passed the attention checks. After excluding participants who failed the comprehension check, the final sample included 1,177 participants (99.6% self-identified females, $M_{age}$ = 42.7 years old). The study employed a 2 (harmful comment moderation: absent vs. present) × 2 (brand transparency: opaque vs. transparent) between-subjects design. The design closely followed that of Study 4A (i.e., using brand attack comments), except that transparency was manipulated by showing participants the brand's social media account profile. In the Brand Opaque condition, this profile simply indicated that it was the brand's official social media account. In the Brand Transparent condition, the profile included a section titled "Community Policy". When comment moderation was present, this section stated that the brand values respectful conversations and therefore hides or removes comments containing harmful language. When harmful comment moderation was absent, it stated that the brand values open conversations and therefore does not moderate any comments—see Figure 8 for exact stimuli. This wording was designed to reflect how brands typically communicate moderation decisions in practice: rather than simply stating that comments are hidden or removed, they tend to provide a rationale for their moderation policies.

For realism, the layout and design of the stimuli were modeled after Instagram's user interface. After viewing the account profile (Figure 8a-c), brand's sponsored post (Figure 8d), and its comments corresponding to the assigned conditions on the same page, participants rated their intentions to click on the ad to purchase the product and their perceived brand trust ($\alpha$ = .92) on 100-point scales.

**Figure 8**. Stimuli Used in Study 5

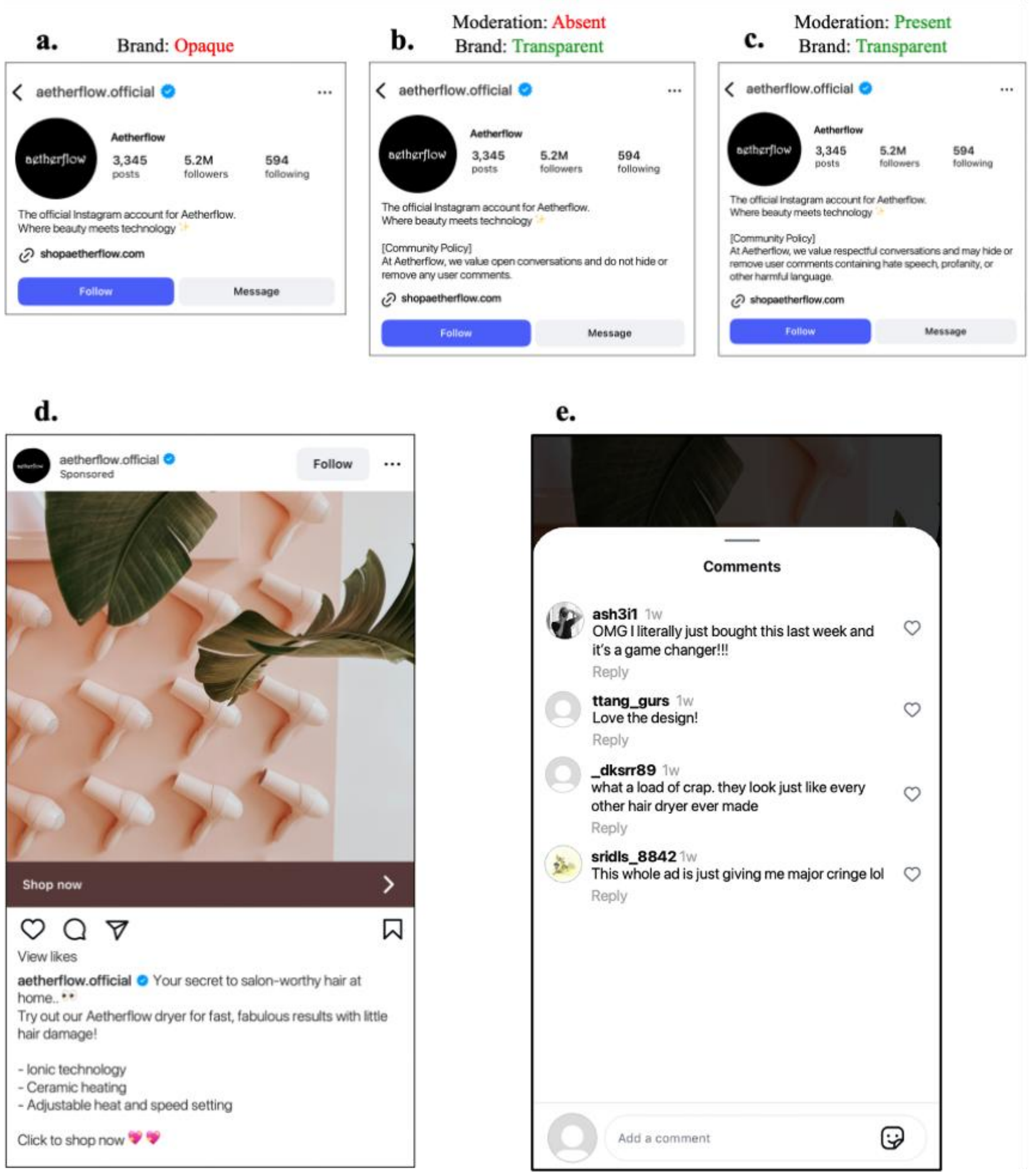

Notes: Figures 8a-c illustrates the brand's account profile across conditions. Figure 8d illustrates the sponsored post, which was shown to all participants. Figure 8e shows an example stimulus used in the comment moderation absent conditions. Brand engagement with positive comments was held constant across all conditions. Given that Study 3 demonstrated that brand engagement had minimal impact on purchase intentions, this design choice was to ensure ecological validity:

without engagement, the comment section would have appeared either blank or consisted only of harmful comments.

### *Results*

A 2×2 ANOVA revealed that only the main effect of comment moderation was significant ($F(1, 1173) = 37.46$, $p < .001$, $\eta^2 = 0.030$). Neither the main effect of brand transparency ($F(1, 1173) = 0.24$, $p = .626$, $\eta^2 < 0.001$) nor the interaction effect ($F(1, 1173) = 1.99$, $p = .158$, $\eta^2 = 0.002$) reached significance. Brand transparency did not remove the positive effect of moderation (H5): participants reported higher purchase intentions when harmful comments were hidden ($M = 57.2$, $SD = 28.0$) than when they were visible ($M = 47.3$, $SD = 27.0$), replicating the harmful comment moderation effect.

Consistent with our preregistration, we conducted pairwise comparisons as a follow up despite the nonsignificant interaction. As in prior studies, when the brand was opaque, participants reported higher purchase intentions when comment moderation was present ($M = 57.9$, $SD = 28.5$) than absent ($M = 45.8$, $SD = 27.5$; $t(1173) = -5.32$, $p < .001$, $d = 0.43$). Importantly, this positive effect of comment moderation persisted even when the brand was transparent: purchase intentions were again higher when comment moderation was present ($M = 56.4$, $SD = 27.5$) than absent ($M = 48.9$, $SD = 26.4$; $t(1173) = -3.33$, $p < .001$, $d = 0.28$). See Figure 9 for results of Study 5.

**Figure 9.** Results of Study 5 (Purchase Intent)

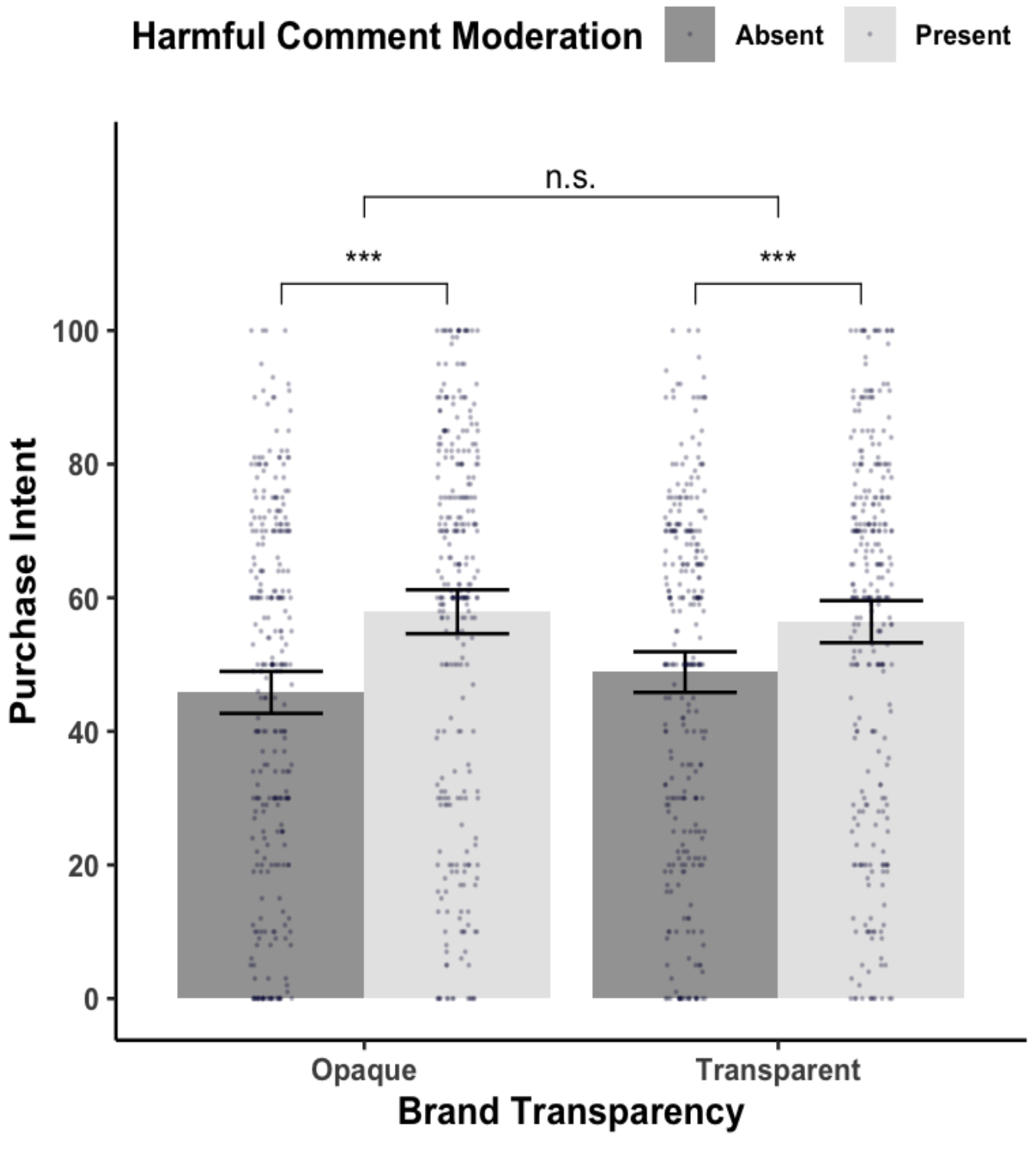

Notes: * $p < .05$, ** $p < .01$, *** $p < .001$

*Perceptions of Brand Trust*

Next, we examined perceived brand trust as the dependent variable. Once again, a 2×2 ANOVA revealed a main effect of harmful comment moderation ($F(1, 1173) = 29.97$, $p < .001$, $\eta^2 = 0.020$). We found a significant main effect of brand transparency ($F(1, 1173) = 4.56$, $p = .033$, $\eta^2 = 0.004$), but importantly, no interaction effect ($F(1, 1173) = 1.42$, $p = .234$, $\eta^2 = 0.001$). Participants perceived the brand as more trustworthy when moderation was present ($M = 59.5$, $SD = 20.2$) than absent ($M = 53.2$, $SD = 19.5$).

*Mediation Analysis*

Given that only the main effect of harmful comment moderation was significant, we conducted a mediation analysis using PROCESS model 4 (Hayes 2017) with 10,000 bootstrapped samples. Harmful comment moderation was included as the independent variable, purchase intentions as the dependent variable, and brand trust as the mediator. As expected, the positive effect of harmful comment moderation on purchase intentions was mediated via perceived levels of brand trust (indirect effect = 6.35, 95% BootCI = [4.07, 8.65]).

***Discussion***

Study 5 shows that hiding harmful comments continues to increase both purchase intentions and brand trust even when the brand is voluntarily transparent about its moderation practices, supporting H5. A Web Appendix study (N = 1,190, Cloud Research) replicates this pattern using universally harmful comments (as in Study 4B): the positive effect of moderation on purchase intentions ($F(1, 1186) = 17.03$, $p < .001$, $\eta^2 = 0.01$) persisted regardless of brand transparency.

## General Discussion

The present research examined whether and when automated comment management—specifically, AI-assisted moderation and engagement—impacts the effectiveness of social media advertising. Across two large-scale field experiments conducted on Trustpilot, Instagram, and Facebook (Studies 1–2), and three preregistered online experiments (Studies 3–5), we find converging evidence that automated comment management improves ad performance. In the field, activating automated comment management improves real advertising outcomes such as consumers' likelihood of completing website registrations (Study 1) and return on ad spend (ROAS; Study 2). Follow-up experiments clarify when and why these benefits arise. Study 3 shows that performance gains from automated comment management are primarily driven by moderating harmful comments (rather than by the brand's engagement with comments). Studies 4A and 4B demonstrate that hiding harmful comments increases purchase intentions by improving perceived brand trust. This benefit reverses when the platform transparently discloses the brand's moderation practices, but only when the hidden harmful comments are directed at the brand—when the hidden comments are universally harmful, then we do not find evidence for a moderating effect of brand transparency. In contrast, Study 5 shows that brand transparency about moderation does not undermine the benefit of moderation, regardless of harmful comment type. Together, these findings show that management of the social context surrounding an ad, namely, the presence or absence of harmful user comments, is a meaningful intervention toward shaping consumer responses and bottom-line advertising outcomes—and there is an interplay between these actions and transparency of this fact by the platform.

***Theoretical contributions***

Our findings contribute to research on contextual effects in marketing (e.g., Aribarg and Schwartz 2020; Goldfarb and Tucker 2011; Shamdasani, Stanaland and Tan 2001; Stipp 2018) and, more specifically, the literature on contextual effects in social media advertising (e.g., Grewal, Stephen and Vana 2025; Lee, Kim and Lim 2021). These studies have focused on surrounding posts or feed content that occur socially and organically, which are outside of the brand's control. In contrast, we focus on user-generated comments on the brand's ads, which are managerially actionable. We find that, by moderating harmful comments (versus leaving them as is), brands can prevent a deterioration in brand trust and thereby increase ad performance.

We also extend the transparency literature (e.g., Eisend et al. 2020; Newman, Howlett and Burton 2016; Wojdynski and Evans 2016), which has shown that transparency can backfire in contexts like disclosures of targeting in advertising (e.g., Boerman, Willemsen and Van Der Aa 2017; Kim, Barasz and John 2019), although in some cases voluntary disclosure can enhance perceived authenticity (e.g., Semaan, Kocher and Gould 2018). Analogous to the first type of result, we find that platform transparency can backfire in that moderating harmful comments decreases purchase intentions—although only in cases where the brand is hiding harmful comments directed at it, rather than universally harmful comments.

Practically speaking, we substantially increase marketing relevance by moving beyond studies of contextual effects that exclusively measure intentions and/or attitudes (e.g., Grewal, Stephen and Vana 2025; Johnson, Voorhees and Khodakarami 2023) within hypothetical scenarios (e.g., Lee, Kim and Lim 2021) to testing real ad performance of various brands on various actual social media sites. We also test how comment moderation is impacted by the practical marketing action of brand transparency of one's moderation practices.

***Managerial implications***

Our findings provide several clear takeaways for marketing managers. First, we suggest that marketers should not only spend on social media advertising itself, but also on managing the comment environment beneath their ads. Today, this is increasingly doable 24/7 at scale, thanks to AI-powered moderation tools. The fact that only around 5% of brands use these sorts of tools suggests that managers are not aware of them (e.g., the second author has found in executive education teaching that executives are often happy and surprised to learn of tools like BrandBastion). It is also possible that the impact of organic comments is under-estimated. For instance, in post-study feedback, one online participant wrote "I just ignore trolls in comments, as I imagine most people do." Yet our results demonstrate that even when harmful comments are not directed at the brand or the ad, and should be widely recognized as unwarranted, their mere presence still degrades brand trust and advertising performance.

Second, and relatedly, our results highlight the need for greater coordination between paid media and community management within firms. In many organizations, social media advertising is managed by paid media teams that focus on spend allocation and measurable performance metrics, whereas comment and community management are handled separately. Our findings suggest that this separation is suboptimal. Because organic user comments directly influence bottom-line advertising outcomes, comment moderation tactics should be viewed as a core component of advertising performance management rather than siloed as a pure reputational concern.

Third, the results suggest that a one-size-fits-all moderation approach across social media platforms is not advisable, especially if the brand moderates brand attacks. Rather, marketers will want to reconsider whether to moderate brand attacks on platforms that have transparency policies. However, we note that in practice it might be that consumers discover hidden comments less often than we manipulate in our studies, depending on how salient platforms make these hidden comments.

Fourth, our results suggest that brands need not be as concerned about voluntary brand transparency, at least insofar as ad performance is concerned. Thus, if there are other good reasons for undertaking brand transparency (e.g., to signal brand values to the community), marketers should be able to take this action without suffering in their advertising performance (as proxied by our measure of purchase intentions).

Fifth, our findings have implications for social media platforms, who need to balance trust in the platform with not scaring away branded advertisers. While it seems intuitive that platform transparency would increase platform trust, we find that it may actually decrease platform trust in some cases. Ironically, this might be because transparency reveals to consumers that platforms are complicit in enabling brands to deceive them into having more positive impressions of the brand's ads than they think brands deserve. Of course, this also raises broader ethical issues around platform governance and deceiving consumers, who may not otherwise have any opportunity to know when they are reading a feed that is a real reflection of user engagement versus curated by the brand. If one believes consumers have the right to know which is which, then platforms arguably have a duty to provide transparency regardless of whether doing so hurts consumers' attitudes toward them. These tensions require an ethical framework that balances consumer, brand, and platform interests.

***Limitations***

Our lab experiments (Studies 3–5) relied on a fictitious hair dryer advertisement for female participants. While we have no theoretical reason to expect the effects of automated comment management are limited to this product category or demographic group, these design choices may constrain generalizability. Future research should replicate these findings across product categories that vary in involvement, risk, and symbolic meaning, as well as across more diverse consumer populations.

Moreover, our transparency manipulations (Studies 4 and 5) involved explicit disclosures of comment moderation. In practice, both the salience of moderation and implementation of transparency policies can vary depending on platform design and contexts. For instance, brands may disclose comment moderation tactics as part of their community guidelines on their profile page (as in our studies), yet users who encounter the advertisement directly may not visit the profile or recall those guidelines at the time of exposure. Similarly, platform-driven disclosures may also be less apparent in practice: for example, TikTok users are only told whether a post has hidden comments after scrolling through the entire comment section, and must click an option to view comments that have been hidden. Thus, real-world transparency cues may be less salient or attended to than in our experimental settings, implying that comment moderation may be less risky to brands in practice. Future research could incorporate attention-tracking measures or experimentally vary the prominence and placement of transparency disclosures to better understand how consumers process these cues in practice.